\documentclass[12pt]{article}

\usepackage{graphicx}
\usepackage{epstopdf}

\newcommand{\beq}{\begin{equation}}
\newcommand{\eeq}{\end{equation}}
\newcommand{\beqa}{\begin{eqnarray}}
\newcommand{\eeqa}{\end{eqnarray}}
\newcommand{\beqar}{\begin{eqnarray*}}
\newcommand{\eeqar}{\end{eqnarray*}}

\begin{document}
\thispagestyle{empty}

\hfill{\sc UG-FT-237/08}

\vspace*{-2mm}
\hfill{\sc CAFPE-107/08}

\vspace{32pt}
\begin{center}

\textbf{\Large Cosmic-ray knee and diffuse 
$\gamma$, $e^+$ and  $\overline p$ fluxes from\\
collisions of cosmic rays with dark matter} 
\vspace{40pt}

Manuel Masip$^{1}$, Iacopo Mastromatteo$^{1,2}$
\vspace{12pt}

\textit{
$^{1}$CAFPE and Departamento de F{\'\i}sica Te\'orica y del
Cosmos}\\ 
\textit{Universidad de Granada, E-18071 Granada, Spain}\\
\vspace{8pt}
\textit{$^{2}$Dipartimento di Fisica Teorica}\\ 
\textit{Universita degli Studi di Trieste, I-34014 Trieste, Italy}\\
\vspace{16pt}
\texttt{masip@ugr.es, iacopomas@infis.univ.trieste.it}
\end{center}

\vspace{40pt}

\date{\today}

\begin{abstract}

In models with extra dimensions the fundamental scale
of gravity $M_D$ could be of order TeV. In that
case the interaction cross section 
between a cosmic proton of energy $E$ and a dark matter 
particle $\chi$ will grow fast with $E$ for 
center of mass energies $\sqrt{2m_\chi E}$ above 
$M_D$, and it could reach 1 mbarn 
at $E\approx 10^9$ GeV. We show that these gravity-mediated 
processes would {\it break} the proton and produce a diffuse 
flux of particles/antiparticles, while {\it boosting} 
$\chi$ with a fraction of the initial proton energy. 
We find that the expected cross sections and dark matter 
densities are not enough to produce an observable 
asymmetry in the flux of the most energetic 
(extragalactic) cosmic rays.
However, 
we propose that unsuppressed TeV interactions may be 
the origin of the {\it knee} observed in the spectrum 
of galactic cosmic rays. The knee would appear at the energy 
threshold for the interaction of dark matter particles
with cosmic protons trapped in the galaxy 
by $\mu$G magnetic fields, and it would imply a well 
defined flux of secondary antiparticles and TeV gamma rays.

\end{abstract}

                            

\newpage

\section{Introduction}

The gravitational interaction between two elementary particles 
is much weaker than the strong or the electroweak ones at
all the energies and distances explored until now in collider 
experiments. 
This is understood due to the large size of the Planck 
mass $M_P\approx 10^{19}$ GeV (in natural units, 
$M_P\equiv G_N^{-1/2}$) 
compared with the energy $E=\sqrt{s}$ in the collision, and 
suggests that gravity may have been only relevant in the 
scatterings  
at the initial moments of the Big Bang (if the 
temperature was ever close the $M_P$).
In recent years, however,
it has become apparent that in models with extra dimensions
this is no longer necessary: the fundamental scale of
gravity $M_D$ could be much lower than $M_P$, and values 
close to the TeV
scale would be {\it natural} in order to solve the hierarchy 
problem \cite{ArkaniHamed:1998rs}. 
Center of mass energies above the TeV would then 
define a transplanckian regime where gravity dominates over
the other interactions \cite{Banks:1999gd}. 

In particular, strong TeV gravity could affect the interactions
of cosmic rays (mostly protons free or bound in 
nuclei \cite{Yao:2006px}) 
that reach the earth with energies of up to $10^{11}$ GeV. 
Notice that 
the relative effect of the new physics would be most relevant 
in processes with a weak cross section 
within the standard model (SM).
In this paper we will focus on the interactions of ultrahigh 
energy cosmic rays with dark matter particles $\chi$ in our 
galactic halo. We will assume that $\chi$ is a 
weakly interacting massive particle (WIMP) of mass 
$m_\chi\approx 100$ GeV, 
although $m_\chi$ could go from 10 MeV to 10 TeV if its 
interaction strength goes from gravitational to 
strong \cite{Feng:2008ya}.
Being a WIMP, an anomalous $p$--$\chi$ interaction rate 
at high energies would be a clear signal 
of nonstandard physics. In addition,
due to its possible large mass, 
the center of mass energy $\sqrt{s}=\sqrt{2 m_\chi E}$ goes 
above the threshold $M_D$ for cosmic rays of 
$E\approx 10^5$ GeV, a region in the spectrum where the 
flux is still sizeable.

First we briefly describe the gravity-mediated 
$p$--$\chi$ interactions 
in the transplanckian regime. Namely, we consider black hole 
production \cite{Dimopoulos:2001ib} and elastic (at the
parton level) scatterings that can be calculated in the
eikonal approximation \cite{Giudice:2001ce}. 
We show that the most important effect
is due to the fact that these interactions  
break the incident proton, producing 
jets that fragment into hadrons and then shower into
stable particles. We use the 
MonteCarlo jet code HERWIG \cite{Corcella:2000bw} to 
determine and parametrize
the flux of photons, neutrinos, electrons and
protons, together with their antiparticles, produced in these 
collisions. Then we 
take a particular dark matter distribution and 
study the probability that an extragalactic
cosmic ray of $E>10^8$ GeV interacts 
with a galactic WIMP in its way to the earth.
This probability, and also the flux 
of secondaries from these interactions, depends
on the galactic longitude, as protons reaching 
the earth from different directions must cross a 
different dark matter column density (depth).
Finally we consider the effect of the gravitational 
interactions on galactic cosmic rays of lower energy.
The crucial difference with the more
energetic ones is that the 
protons of energy below $10^8$ GeV 
are trapped by the $\mu$G magnetic fields 
\cite{Battaner:2000ef} 
present in our galaxy (their Larmor radius is 
contained in a typical cell of magnetic field). 
As a consequence, the dark matter depth that they face
grows with time as they diffuse from the center of
the galaxy, and a significant fraction of them 
may interact before reaching the earth.

\section{Interactions at transplanckian energies}

The possibility to produce black holes (BHs) in the collision
of two particles at $\sqrt{s}> M_D$ has been 
extensively entertained in the literature \cite{Dimopoulos:2001ib}. 
Basically, one
expects that for impact parameters smaller than the
horizon $r_H$ of the system they collapse into a BH of mass 
$M\approx \sqrt{s}$. Here we will assume
that there are $n$ flat 
extra dimensions of common length where gravity propagates, 
that all matter fields are trapped on a 4-dimensional brane,
and that $r_{H}$ is
just the higher dimensional Schwarzschild radius,
\beq
r_{H}=\left({2^n\pi^{n-3\over 2}\Gamma\left({n+3\over 2}\right)
\over n+2}\right)^{1\over n+1}
\left({M\over M_D}\right)^{1\over n+1}
     {1\over M_D}\;.
\label{rh}
\eeq
Therefore, for two pointlike particles 
the cross section $\sigma_{BH}=\pi r_H^2$ to produce a BH 
(in Fig.~1) is
of order $1/M_D^2$ and grows like $s^{1/(n+1)}$ with the 
center of mass energy.
\begin{figure}
\begin{center}
\includegraphics[width=0.5\linewidth]{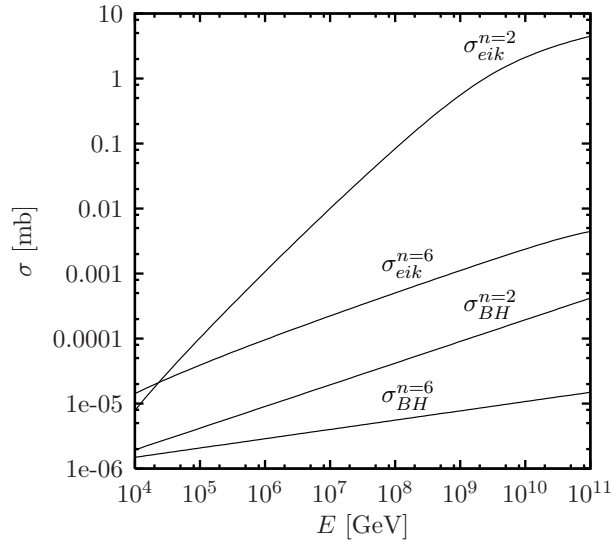}
\end{center}
\caption{Cross section for gravitational  
interactions (eikonal and BH production) in a
collision of two point-like particles for $n=2,6$, $M_D=1$ TeV 
and $m_\chi=200$ GeV.
\label{fig1}}
\end{figure}

Gravity-mediated interactions, however, are also important at 
impact parameters $b$ larger than $r_H$ \cite{Giudice:2001ce}. 
In such processes an 
incident particle of energy $E$ will interact elastically with a
target at rest and will lose a small fraction $y$ of its energy. 
The process can be calculated in the eikonal approximation, that 
provides a resummation of 
ladder and cross-ladder contributions. It has been
shown \cite{Illana:2005pu}
that effects like the dependence on the physics 
at the cutoff $M_D$ or the emission of gravitons during the 
collision are negligible if $\sqrt{s}\gg M_D$.
For two pointlike particles the
eikonal amplitude can be written 
\beq
{\cal A}_{eik}(s,q)=4\pi s b_c^2\; F_n(b_c q)\;, 
\label{eik}
\eeq
where $q=\sqrt{-t}=\sqrt{2ymE}$ is the exchanged transverse 
momentum,
\beq
b_c=\left({(4\pi)^{{n\over 2}-1}\Gamma(n/2)\over 2}\right)^{1\over n}
\left({\sqrt{s}\over M_D}\right)^{2\over n}{1\over M_D}\;,
\label{bc}
\eeq
and the functions 
\beq
F_n(y)=-i 
\int_0^\infty {\rm d}x\;x\; J_0(xy)
\left( e^{ix^{-n}} -1 \right)\;
\label{eq7}
\eeq
are given in \cite{Sessolo:2008ex} in terms of Meijer's G-functions. 
The eikonal cross section grows fast with the energy for low
values of $n$, it goes like $s^{1+4/n}$.
In Fig.~1 we plot the cross sections $\sigma_{BH}$ and $\sigma_{eik}$ 
between two point-like particles for $n=2,6$,
$M_D=1$ TeV, $m_\chi=200$ GeV and values of the incident energy
of up to $10^{11}$ GeV. We have required that the transverse
momentum is $q>1$ GeV, which sets 
a minimum value of $y=(E-E')/E$ (notice that the eikonal amplitude
diverges at $y=0$ for $n\le 2$).
\begin{figure}
\begin{center}
\includegraphics[width=0.5\linewidth]{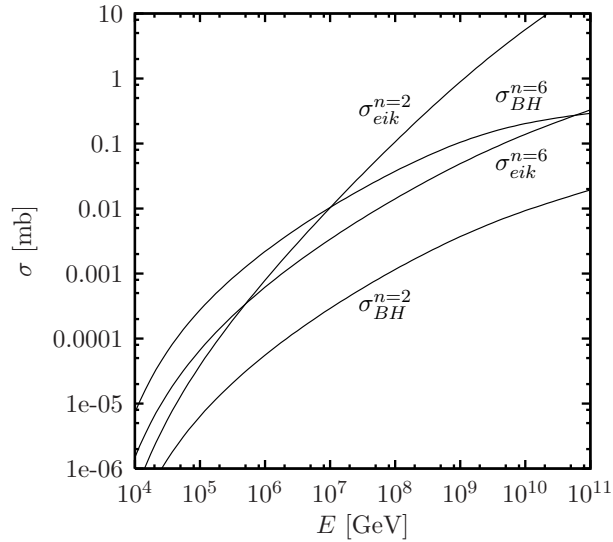}
\end{center}
\caption{Cross section for gravitational interactions in a
$p$--$\chi$ collision for $n=2,6$, $M_D=1$ TeV and $m_\chi=200$ GeV.  
\label{fig2}}
\end{figure}

When a cosmic proton of energy $E$ hits a dark
matter particle $\chi$ initially 
at rest, the collision will be dominated by 
transverse distances ($r_H$ or $b_c$) much smaller than 
the proton radius. This indicates that $\chi$ {\it sees} the 
proton structure and interacts with 
a parton carrying a fraction $x$ of the proton momentum.
The transplanckian regime requires then 
$\sqrt{\tilde s}=\sqrt{2xm_\chi E}>M_D$. In Fig.~2 we plot
$\sigma^{p\chi}_{BH}$ and $\sigma^{p\chi}_{eik}$ for the same
choice of parameters as in Fig.~1. 
We have used the CTEQ6M \cite{Pumplin:2002vw} parton 
distribution functions. 

If the parton (q) and $\chi$
form a mini BH (in Fig.~3--left), 
after the collision we will have two {\it jets}: the 
BH, of mass $\sqrt{2xm_\chi E}$, energy $xE$ and 
the color of the parton involved in the collision,
plus the proton remnant (qq, the spectator partons), 
with the opposite color and energy $(1-x)E$. 

On the other hand, if the collision of q with $\chi$ 
is elastic (in Fig.~3--right), they will result into a boosted 
dark matter particle of energy $xyE$ plus two jets (indicated with
dotted cones in Fig.~3): the one defined
by the scattering parton, with energy $x(1-y)E$ and transverse
momentum $p_T=\sqrt{2xym_\chi E}$, and a proton
remnant qq of opposite color and energy $(1-x)E$.
\begin{figure}
\begin{center}
\includegraphics[width=1.0\linewidth]{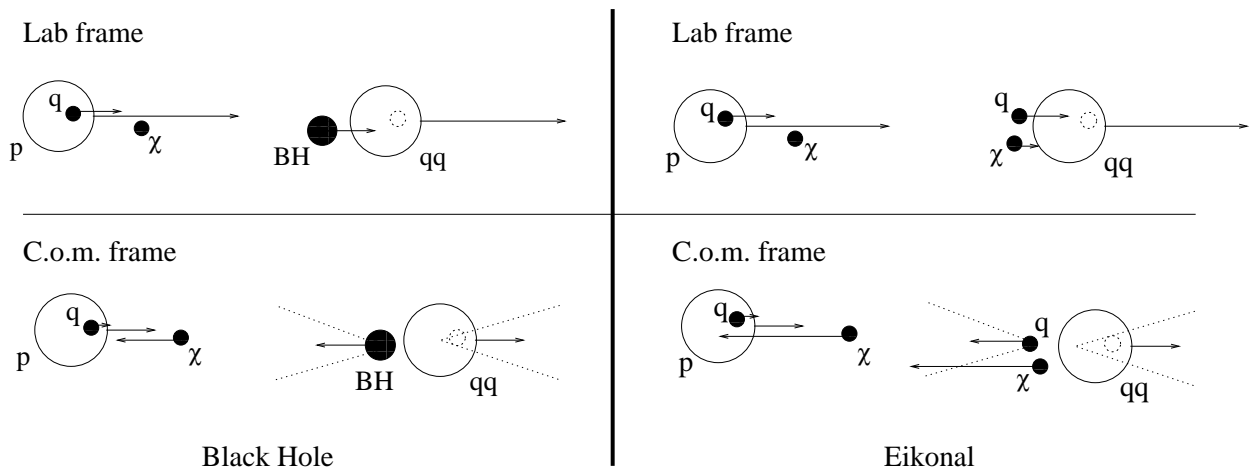}
\end{center}
\caption{Scheme of a $p$--$\chi$ collision giving a BH (left) or
an eikonal process (right) in the lab (upper) and in the c.o.m. (lower)
frames. In a BH process the final states are
the proton remnant (qq) and the BH, whereas in the eikonal process
we have the scattering parton (q), the proton remnant (qq) and the 
dark matter particle ($\chi$). The center of mass frame
in the eikonal process refers to the two jets (q and qq), not
the $p$--$\chi$ system.
\label{fig3}}
\end{figure}

\section{Fragmentation and decay into stable species}

To understand how the system evolves after the gravitational
interaction, it is convenient to study the process in the 
center of mass frame of the two final jets.
Let us start discussing 
the {\it elastic} (at the parton level) scattering of 
a dark matter particle $\chi$ 
with a parton q. The process will break the incident proton,
since the typical transverse momentum exchanged 
is larger than 1 GeV.
For $x,y\ll 1$ we go to that frame with a boost of
\beq
\gamma \approx \sqrt{E\over 2 y m_\chi}\;.
\label{gammaeik}
\eeq
There the energy of the two jets becomes 
\beq
E_q = E_{qq} \approx \sqrt{y m_\chi E\over 2}\;.
\label{ejets1}
\eeq
The scattering parton and the proton remnant will then emit 
gluons and quarks, fragment into
hadrons, and shower into stable species. We evaluate this process
using HERWIG \cite{Corcella:2000bw}. In this center of mass fram
we find that 

{\it (i)} The scattering parton and the proton remnant define
jets giving a very similar spectrum of stable particles.
This spectrum is only mildly sensitive to 
the fact that the parton
may be a quark or a gluon and the proton remnant may be, 
correspondingly, a diquark 
(color antitriplet) or a triquark (color octet). 
When the scattering parton is a gluon 
the jets tend to give a larger number
of stable particles of smaller energy than when it is a
quark. In Appendix A we include two Tables containing
the spectrum (averaged over 100 HERWIG runs)  
of 100 GeV  quark-diquark jets 
and gluon-triquark jets.

{\it (ii)} In this frame the final spectrum of stable particles 
is dominated by 
energies around 1 GeV, almost independently of the energy
of the parton starting the shower. In Table 3 (also in the
appendix) we include the spectrum from a 10 GeV di-jet.

{\it (iii)} The stable species (particle plus antiparticles) 
are produced with a frequency $f_i$ that is mostly independent
of the energy or the nature of the two jets. We obtain an
approximate 
55\% of neutrinos, a 20\% of photons, a 20\% of electrons, and
a 5\% of protons.

We will then parametrize these fluxes in terms of functions
$g_i(E,E_{jet})$ that indicate the number of particles of the
species $i$ coming from a jet of energy $E_{jet}$:
\beq
N_i=f_i N = \int {\rm d}E\; g_i (E;E_{jet})\;.
\eeq
For photons, electrons and neutrinos we will use the ansatz
\beq
g_i(E;E_{jet}) = \left\{
\begin{array}{l l} 
\displaystyle
{f_i N\over \Lambda} \left( \frac{\beta - 1}{\beta} 
\right) & E < \Lambda\;; \\
\displaystyle
\frac{f_i N}{\Lambda} \left( \frac{\beta - 1}{\beta} 
\right) \left( \frac{E}{\Lambda} \right)^{-\beta} & 
E_{jet} > E > \Lambda\;,
\end{array} \right. 
\label{gi}
\eeq
whereas for the proton 
\beq
g_p(E;E_{jet}) = {f_p\over f_i}\; g_i(E-m_p;E_{jet}) \;.
\eeq
In these expressions we take 
$\Lambda=0.2$ GeV and fix $\beta$ by energy conservation:
\beq
E_{jet} \approx N\Lambda\;\frac{\beta -1}{2 (\beta - 2)}+f_p N m_p\;.
\eeq
We plot in Fig.~4 the number of particles 
$N$ and the parameter $\beta$ that we obtain for the two types of
jets (quark-diquark and gluon-triquark) 
of energy between $10$ and $10^4$ GeV.
\begin{figure}
\begin{center}
\begin{tabular}{cc}
\includegraphics[width=0.5\linewidth]{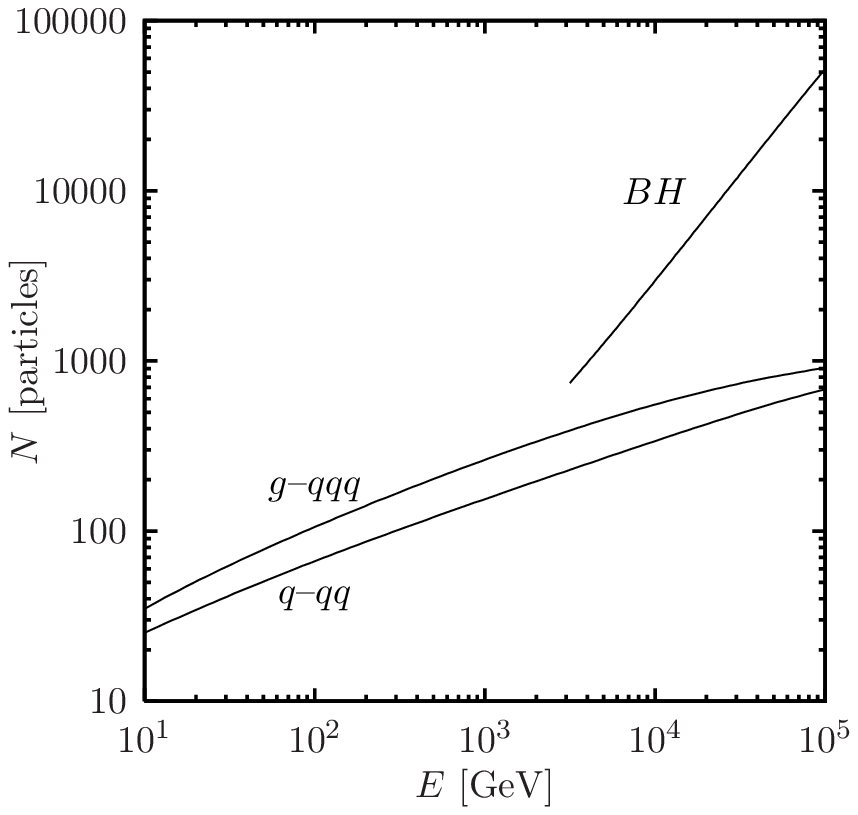} &
\includegraphics[width=0.5\linewidth]{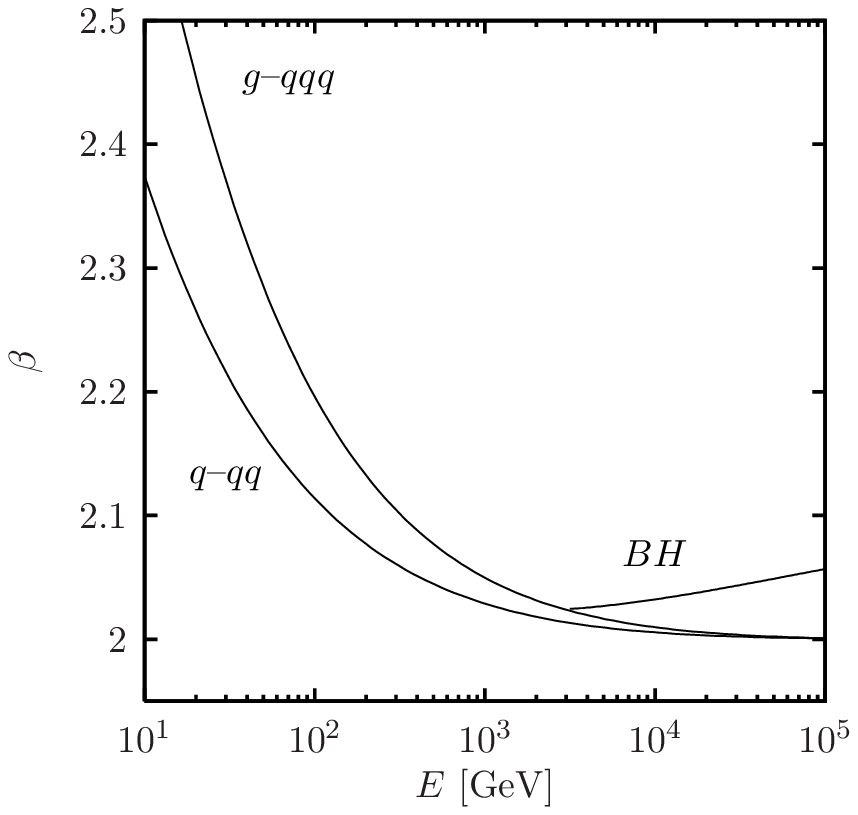} 
\end{tabular}
\end{center}
\caption{Value of the total number of stable particles $N$ (left) 
and of the spectral index $\beta$ (right) used in the parametrization
of quark-diquark jets and gluon-triquark jets of energy $E_{jet}=E$ and
of BHs of mass $M=E$.
\label{fig4}}
\end{figure}

In processes where a parton and $\chi$ collapse into a BH
(see Fig.~1--left) the situation is not too diferent. 
In the lab frame the mass and energy of the BH produced
are $M=\sqrt{2 x m_\chi E}$ and $\approx xE$, respectively,
whereas the proton remnant has an energy $(1-x)E$.
We go to the center of mass frame with a boost  
\beq
\gamma \approx \sqrt{E\over 2 m_\chi}\;,
\label{gammaBH}
\eeq
which leaves
both {\it jets} with oposite momenta of order 
\beq
E_{BH} = E_{qq} \approx \sqrt{m_\chi E\over 2}\;.
\label{ejets2}
\eeq
In this frame the proton remnant will produce 
a jet similar to the ones described above. The BH, however,
follows a different path, as it 
emits radiation \cite{Hawking:1974rv}
and evaporates in a typical 
timescale much shorter than $\Lambda_{QCD}^{-1}$.
The Hawking evaporation of these BHs has been analyzed
in \cite{Draggiotis:2008jz}. 
Very briefly, the results in there can be summarized
as follows.

{\it (i)} The spectrum of stable particles resulting from
a mini BH is very similar to the one obtained from the 
quark and gluon jets described before. It 
consists of an approximate 55\% of neutrinos, 20\% of 
photons, 20\% of electrons and 5\% of protons. 
For $M_D=1$ TeV and $n=2$, gravitons account for a 1\% of 
the total energy emitted by a BH of $M=10^4$ GeV. 
If there are $n=6$ extra dimensions graviton emission 
grows to the 15\%.

{\it (ii)} In the BH rest frame the spectrum of stable
particles is dominated by energies below 1 GeV. In the 
lab frame this energy is boosted by a factor of 
$\gamma_{BH}\approx \sqrt{xE/(2m_\chi)}$.

Therefore, we will also use Eq.~(\ref{gi}) to parametrize
the spectrum of stable particles resulting 
from the evaporation of
a BH in its rest frame. We substitute $E_{jet}$ for
the BH mass $M$, discount from $M$ the energy taken by
the emitted gravitons, and include in Fig.~4 the total
number of stable particles and the spectral index $\beta$
for light BHs of mass between 1 and 10 TeV.

\section{Secondary flux from extragalactic cosmic rays}

In their way to the earth cosmic rays cross a medium full
of dark matter particles. 
If the gravitational interaction becomes
strong above the TeV, a fraction of them
will collide,
inducing a flux of secondary particles that might be
observable at satellite \cite{pamela,Gehrels:1999ri} or ground
based \cite{Abdo:2007ad} experiments. The probability that 
a cosmic ray interacts will depend on the cross section 
$\sigma$ and on the column density 
(depth $x$) of dark matter along its trajectory:
\beq
p(x)\approx {\sigma\; x\over m_\chi} \;,
\label{prob}
\eeq
where
\beq
x=\int \rho\; {\rm d}l \;
\label{depth}
\eeq
and $\rho$ is the dark matter density at each point
of the trajectory.
Throughout the paper 
we will assume $10^{69}$ GeV of galactic 
dark matter distributed in a sphere of 200 kpc with the
density profile \cite{Navarro:1995iw}
\beq
\rho(r)\approx {\rho_0\over {\left( {r\over R}\right) 
\left( 1+ {r\over R}\right)^2}}\;,
\label{eq3}
\eeq
where $R=20$ kpc (we are at 8 kpc from the center, 
$1\; {\rm kpc}=3\times 10^{19}$ m). 
This means that the depth $x$ of dark matter from the
earth to the border of our galaxy goes from 
0.01 g/cm$^2$ for a galactic latitude $\theta=180^o$ 
to 0.17 g/cm$^2$ for a trajectory crossing the galactic
center with $\theta=5^o$ (notice that the depth
diverges at $\theta=0$). For the dark matter particle
we will take $m_\chi=200$ GeV, although this value
could oscillate within a wide range depending on its
anihilation and coanihilation cross sections. 

In this section we will focus 
on cosmic rays of extragalactic origin and 
energy above $10^8$ GeV. Since 
their interaction probability grows with
the dark matter depth, 
it may change in one order of magnitude 
depending on the angle $\theta$ of approach to 
the earth. If this probability were sizeable, it 
would deplete the flux of ultrahigh 
energy protons and create an observable asymmetry. 
Taking $\theta=90^o$ ($x\approx 0.02$ g/cm$^2$ from
the border of the galaxy) and 
$\sigma=\sigma_{BH}+\sigma_{eik}$, we obtain 
interaction probabilities
that go from $1.2\times 10^{-3}$ 
for a proton of $E=10^{11}$ GeV to
$0.7\times 10^{-5}$ for $E=10^{8}$ GeV 
(with $n=2$ and $M_D=1$ TeV). 
This probability is dominated by $\sigma_{eik}$, 
where we only include exchanged momenta large enough 
to {\it break} the incident proton ($q_T>1$ GeV).
The one in a thousand depletion in the flux from 
interactions within
the galaxy seems too small to be observable, 
although the effect would be larger for fluxes from 
distant sources crossing regions of large dark
matter density.

To obtain the total flux of stable particles reaching
the earth we convolute the two processes under study
(BH production and eikonal collisions) with the primary
proton flux. We first define the event at the parton 
level. Then we go to the center of mass frame of the
two jets (proton remnant--scattering parton {\it or} 
proton remnant--BH),
where we get the spectrum of stable 
species using the parametrization discussed
in the previous section (the BH evaporation is obtained 
in the BH rest frame). Finally, we boost these 
spectra to the lab frame. We also calculate the 
energy of the dark matter particle $\chi$ after
the eikonal scattering with the proton.

\begin{figure}
\begin{center}
\includegraphics[width=0.5\linewidth]{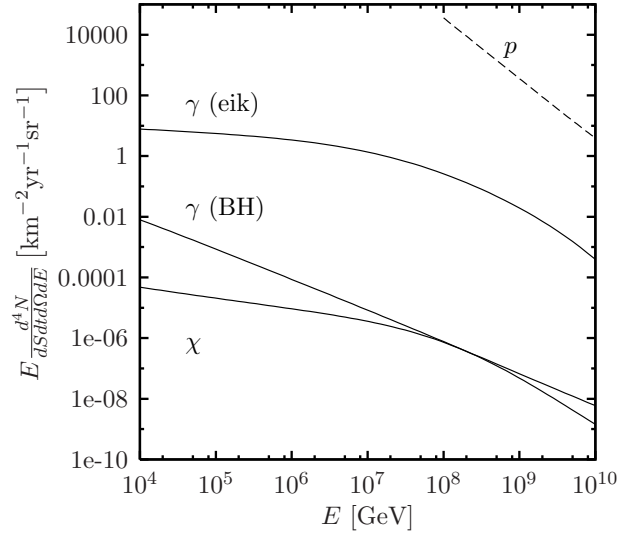}
\end{center}
\caption{Photon flux from (eikonal and BH) collisions of 
extragalactic protons
of $E>10^8$ GeV with dark matter particles 
for $n=2$, $M_D=1$ TeV and $m_\chi=200$ GeV.
We take a galactic latitude of $\theta=90^o$ ({\it i.e.},
a dark matter depth of 0.02 g/cm$^2$). We also plot the
flux of dark matter particles accelerated by the eikonal
collision.
\label{fig5}}
\end{figure}
In Fig.~5 we plot the flux of
photons and boosted dark matter particles
for $n=2$ and $M_D=1$ TeV from interactions
of cosmic rays of extragalactic origin and energy
above $10^8$ GeV. The 
flux of the other species has basically the same 
spectrum (up to propagation effects) as the 
flux of photons.
It is a factor of 2.7 larger for neutrinos, 
similar for electrons, and a factor of 4 smaller
protons. These fluxes include the same 
number of particles and antiparticles of any species.
In Fig.~5 we separate the photons produced 
in eikonal and in BH events. 
It is apparent that the dominant source of
secondaries is the elastic (eikonal) scattering 
of the partons in the proton with dark matter particles.

\section{Flux from galactic cosmic rays}

The effect of an anomalous proton--dark matter interaction 
rate could be most relevant for the cosmic rays of 
$E<10^8$ GeV of predominantly galactic origin. The reason 
is that a large fraction of them will be trapped by the random 
magnetic fields of order $\mu$G present in our galaxy (the 
Larmor radius of their trajectory, around 0.1 kpc for 
a $10^8$ GeV 
proton, would be contained in a typical cell of magnetic field).
From the moment they are produced, these protons may collide 
with $\chi$, being their interaction length around
$L\approx m_\chi/(\sigma\rho)$. Now, a remarkable feature 
of the gravitational interaction that we are considering is
that it grows fast above transplanckian energies, much
faster than, for example, $Z$ boson exchange. 
In particular, if $m_\chi=200$ GeV, $n=2$ and 
$M_D=5$ TeV, then at proton energies around $10^{5}$ GeV the 
gravitational interaction would be {\it weak} (with a pbarn 
cross section), whereas at at $10^{7}$ it would be four orders 
of magnitude larger (see Fig.~6).
\begin{figure}
\begin{center}
\includegraphics[width=0.5\linewidth]{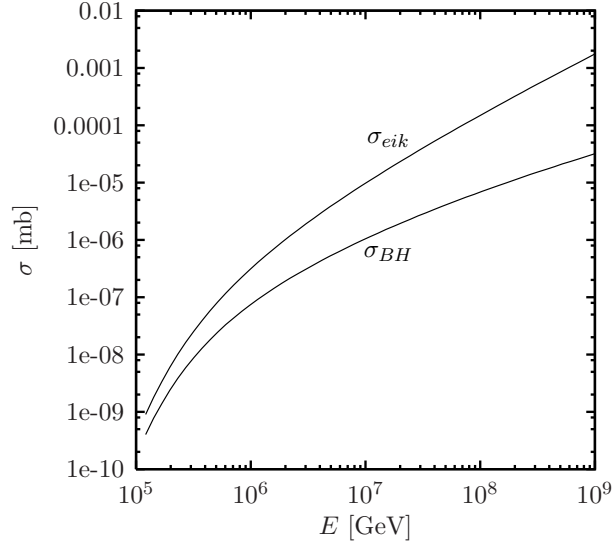}
\end{center}
\caption{Cross section for a gravitational interaction in a
$p\chi$ collision for $n=2$, $M_D=5$ TeV and $m_\chi=200$ GeV.
\label{fig6}}
\end{figure}
An increased collision rate of protons above
the transplanckian threshold would reduce their abundance,
producing an effect in the flux reaching the earth 
that could explain the {\it knee} 
observed at $10^{6}$ GeV.

The cosmic ray flux up to energies around 
$10^{6}$ GeV is 
\beq
\frac{{\rm d}\Phi_{N}}{{\rm d}E}\approx 
1.8 \; E^{-2.7}\; {\rm {nucleons\over cm^2\;s\;sr\;GeV}}\;.
\label{pcr}
\eeq
If gravitational interactions were responsible for
the change in the 
diferential spectral index from 2.7 to 3, then there 
would be a flux of secondary particles that could be
readily estimated. Let us assume that, on absence of
gravitational interactions, the flux in
(\ref{pcr}) would have extended 
up to $10^{8}$ GeV. This means that the flux 
\beq
\Phi_{N}\approx \int_{10^6{\rm\; GeV}}^{10^8{\rm\; GeV}}
{\rm d}E\; 1.8 \; (E^{-2.7}-10^{1.8} E^{-3})
\; {\rm {nucleons\over cm^2\;s\;sr}}\label{en}
\eeq
had been {\it processed} by these interacions into 
secondary particles of less energy. We plot in Fig.~7
the proton and gamma-ray fluxes 
for $M_D=5$ TeV, $m_\chi=200$ GeV and
$n=6$ (the flux of neutrinos would be a factor of 2.7 
more abundant than the $\gamma$-ray flux), together with the 
flux of dark matter particles boosted by the 
eikonalized scattering.
\begin{figure}
\begin{center}
\includegraphics[width=0.5\linewidth]{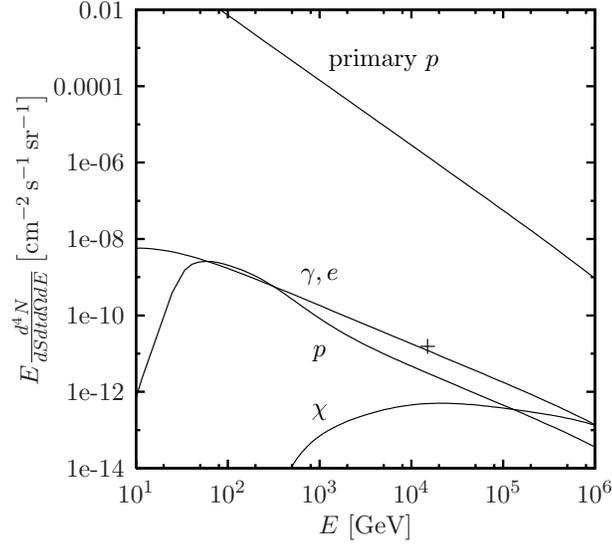}
\end{center}
\caption{Secondary fluxes from $p$--$\chi$ gravitational collisions
 for $n=6$, $M_D=5$ TeV and $m_\chi=200$ GeV. The fluxes include the 
same amount of particles and antiparticles. The point at 15 TeV
 indicates the gamma-ray flux measured by MILAGRO.
\label{fig7}}
\end{figure}
The flux of $e=e^++e^-$ is similar to the 
photon flux, although
the propagation effects (sicroton emission, etc.) that may
distort the spectrum have not been included. 
Recent data from PAMELA \cite{Adriani:2008zr} 
signals an excess in the positron flux 
above 10 GeV, although the contribution that we
find seems to be well below these data. 
We add in the plot the diffuse gamma-ray flux 
measured by MILAGRO \cite{Abdo:2008if} at energies around 15 TeV, 
which seems to indicate an excess versus the expected values from 
some regions in the galactic plane. Previous measurements of this 
flux have always indicated a TeV 
excess \cite{Atkins:2005wu,Prodanovic:2006bq,Casanova:2007cf}, 
while EGRET observations
\cite{Hunger:1997we,Baughman:2007ck} have also pointed 
to a harder spectrum than expected, with
too many gamma rays in the 1--10 GeV region. The contribution
that we find could explain anomalies in the gamma-ray flux
above 10 GeV or in the positron and antiproton fluxes 
above 1 TeV.

In Fig.~8 we give the gamma flux 
\begin{figure}
\begin{center}
\includegraphics[width=0.5\linewidth]{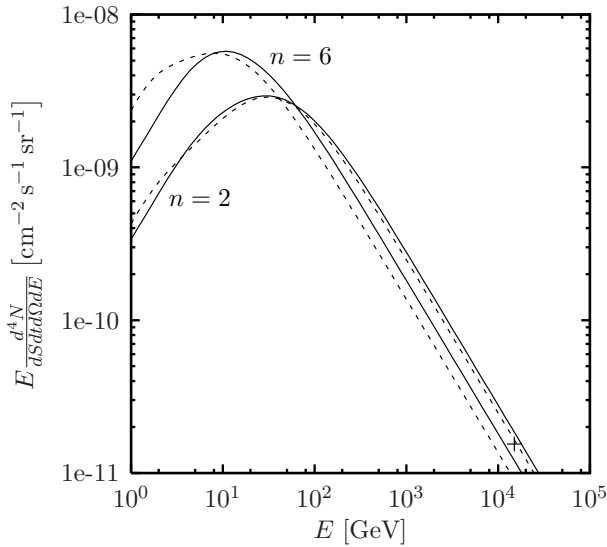}
\end{center}
\caption{Flux of secondary photons from $p$--$\chi$ gravitational 
collisions for $M_D=5$ TeV, $n=2,6$, and $m_\chi=100$ GeV (solid) 
or $m_\chi=500$ GeV (dashes). For $n=2$ 80\% of the flux comes from 
eikonal interactions, whereas for $n=6$ BH production
dominates and is the origin of 85\% of the secondaries. 
\label{fig8}}
\end{figure}
for the different values of the dark matter mass and the 
number of extra dimensions. We observe that the spectrum changes
only slightly with these parameters, while the frequency of
the different species in the flux of secondaries is
model independent. 

\section{Summary and discussion}

Strong gravity at the TeV scale could affect the 
propagation of the most energetic cosmic rays through
the galactic and intergalactic media. In particular,
cosmic protons could interact with the WIMP $\chi$ that 
constitutes the dark matter of our universe.
These interactions would be in addition to the 
standard ones with nucleons in the interstellar medium,
where the effect is negligible \cite{Barrau:2005zb}.
Basically, there are three consequences from 
such processes. 

{\it (i)} Since the interaction breaks the incident
proton, if a sizeable fraction of them 
interacts we would observe a reduced  
flux reaching the earth. At energies around $10^{10}$ GeV
({\it i.e.}, for cosmic rays not trapped by the galactic
magnetic fields)
this reduction would be stronger 
from directions crossing regions with a larger 
dark matter density, whereas at lower energies 
(trapped protons) it increases with the 
{\it length} of the proton trajectory. In both cases 
the suppresion {\it measures} the product of 
the dark matter column density faced by the flux 
in its way to 
the earth and the $p$--$\chi$ cross section, that grows
with the energy of the cosmic ray.

{\it (ii)} The interactions would produce a flux of
secondary particles. This flux includes basically 
the same amount 
of particles and antiparticles, and it grows with the flux 
of primaries (they should have similar angular dependence) and 
with the dark matter depth to reach the earth. Its origin
would be the jets defined by the proton 
remnant (the spectator partons) and the scattering parton
in gravitational interactions with $\chi$, and also 
the BH decay 
products.

{\it (iii)} The third generic effect would be a flux of 
dark matter particles boosted by the 
eikonal interactions with cosmic protons.
This flux, also proportional to the flux of primaries 
and to the dark matter depth,
would only depend on the mass of $\chi$,
not on its electroweak interactions.

For very energetic extragalactic protons, 
we have shown that the gravitational interaction and
the dark matter density are not large 
enough to produce an asymmetry in the cosmic ray
flux reaching the earth, 
as only one in a thousand of the $10^{10}$ GeV protons 
could experience a BH or an eikonal interaction
(an asymmetry would require cross sections or
dark matter densities 100 times larger).
The flux of secondary antiparticles or gamma
rays that we obtain (in Fig.~5) is also negligible.

We have argued, however, that these interactions could have
an observable effect on the less energetic ($E < 10^8$ GeV)
protons of galactic origin. The reason is that a large
fraction of them are trapped by the random 
magnetic fields of order $\mu$G present in our galaxy. 
The dark-matter column density 
that these protons face grows with time, and eventually a gravitational
interaction could {\it break} them before they reach the earth.
We have speculated that this may be
the origin of the knee in the cosmic ray flux at energies
around $10^6$ GeV. If that were the case, then the secondaries
would define a TeV flux of antiprotons, positrons and gamma rays
that could be observable at GLAST \cite{Gehrels:1999ri}, 
MILAGRO \cite{Abdo:2007ad} or PAMELA \cite{pamela}.
Notice that the energy dependence of these fluxes (to be 
measured in the near future) is clearly different from
the one obtained from dark matter anihilation 
\cite{Bringmann:2007nk,Cirelli:2008jk,Barger:2008su,Cholis:2008hb} or
pulsars \cite{Hooper:2008kg}.

A proof that the {\it knee} has anything to do with 
strong TeV gravity or with other new physics producing an
anomalous proton--dark matter interaction rate 
would require an elaborate model for the production of
galactic cosmic rays together with a simulation of 
the effects caused by the galactic magnetic fields on
the propagation of primary and secondary cosmic rays. 
In any case, it is puzzling that the change in the 
spectral index in the flux appears at center of
mass energies $\sqrt{2m_\chi E_{knee}}\approx 10$ TeV,
where the new physics is expected. The {\it knee} 
could be more related to the destruction of cosmic rays 
in collisions with dark matter 
than to the production mechanisms at energies around 
$10^6$ GeV. We think that the experimental observation 
of a flux of 
gamma rays and antiparticles with the spectrum in Fig.~7 
would provide strong support for this hypothesis.

\section*{Acknowledgments}
We would like to thank Eduardo Battaner for very useful 
discussions about galactic magnetic fields.
This work has been supported by MEC of Spain 
(FPA2006-05294) and by
Junta de Andaluc\'\i a (FQM-101 and FQM-437).
I.M. acknowledges a grant from 
C.F. Luciano Fonda (Italy).

\appendix

\section{Stable particles from jets}

In this appendix we include several tables with the 
spectrum of stable species resulting from di-jets
of different nature, in the center of mass of the
two jets. They have been obtained simulating  
elastic collisions with the code
HERWIG, averaging over 100 runs in each case. The
scattering parton may be a quark (q) or a gluon (g), 
whereas the proton remnant would be, correspondingly, 
a diquark (qq) or a triquark (qqq). The spectra
include particles and antiparticles (for $\nu$, $e$ 
and $p$), and for $\nu$ it also includes the three 
neutrino flavours.

\begin{table}
\begin{center}
\begin{tabular}{rl}
\begin{tabular}{r}
from q-jet\\
(100 GeV)
\\ \\ \\ \\ \\ \\ 
from qq-jet
\\
(100 GeV)
\end{tabular}
&
\begin{tabular}{|c|c|c|c|c|c|}
\hline
{Energy [GeV]} & $\nu$ & $e$ & $p$ & $\gamma$ 
& TOTAL 
\\
\hline
$10^{-2}$--$10^{-1.5}$       & 8.8 & 3.3 & 0 & 1.6 & 13.8 
\\
$10^{-1.5}$--$10^{-1}$       & 11.5 & 4.2 & 0 & 3.4 & 19.3
\\
$10^{-1}$--$10^{-0.5}$       & 9.9 & 3.6 & 0 & 4.3 & 17.8 
\\
$10^{-0.5}$--$1$       & 6.3 & 2.1 & 0.74 & 3.4 & 12.6 
\\
$1$--$10^{0.5}$       & 2.3 & 0.84 & 0.94 & 1.7 & 5.8 
\\
$10^{0.5}$--$10$       & 0.55 & 0.21 & 0.53 & 0.54 & 1.8 
\\
\hline
TOTAL & 39.5 & 14.3 & 2.2 & 15.1 & 71.2
\\
\hline
\hline
$10^{-2}$--$10^{-1.5}$       & 7.4 & 5.1 & 0 & 1.3 & 13.8 
\\
$10^{-1.5}$--$10^{-1}$       & 9.8 & 5.5 & 0 & 2.9 & 18.2
\\
$10^{-1}$--$10^{-0.5}$       & 8.9 & 5.2 & 0 & 4.0 & 18.1 
\\
$10^{-0.5}$--$1$       & 4.8 & 2.8 & 0.75 & 2.9 & 11.3 
\\
$1$--$10^{0.5}$       & 1.8 & 0.71 & 0.76 & 1.52 & 4.8 
\\
$10^{0.5}$--$10$       & 0.28 & 0 & 0.50 & 0.64 & 1.4 
\\
$10$--$10^{1.5}$       & 0 & 0 & 0.15 & 0 & 0.15 
\\
\hline
TOTAL & 33.0 & 19.4 & 2.2 & 13.3 & 67.8
\\
\hline
\end{tabular}
\end{tabular}
\end{center}
\caption{
Spectrum of stable species from quark and diquark jets of 100 GeV. 
\label{tab1}}
\end{table}

\begin{table}
\begin{center}
\begin{tabular}{rl}
\begin{tabular}{r}
from g-jet\\
(100 GeV)
\\
\\
\\
\\
\\
\\
from qqq-jet
\\
(100 GeV)
\end{tabular}
&
\begin{tabular}{|c|c|c|c|c|c|}
\hline
{Energy [GeV]} & $\nu$ & $e$ & $p$ & $\gamma$ 
& TOTAL 
\\
\hline
$10^{-2}$--$10^{-1.5}$       & 14.7 & 5.6 & 0 & 2.5 & 22.8 
\\
$10^{-1.5}$--$10^{-1}$       & 18.5 & 6.5 & 0 & 5.9 & 30.9
\\
$10^{-1}$--$10^{-0.5}$       & 16.6 & 5.7 & 0 & 7.2 & 29.6 
\\
$10^{-0.5}$--$1$       & 8.4 & 3.0 & 0.1.5 & 5.0 & 17.8
\\
$1$--$10^{0.5}$       & 2.3 & 0.75 & 1.5 & 2.2 & 6.7 
\\
$10^{0.5}$--$10$       & 0.20 & 0.09 & 0.42 & 0.41 & 1.1 
\\
\hline
TOTAL & 60.6 & 21.7 & 3.4 & 23.2 & 108.9
\\
\hline
\hline
$10^{-2}$--$10^{-1.5}$       & 15.2 & 5.7 & 0 & 2.7 & 23.6 
\\
$10^{-1.5}$--$10^{-1}$       & 18.6 & 6.3 & 0 & 5.9 & 30.7
\\
$10^{-1}$--$10^{-0.5}$       & 16.2 & 5.9 & 0 & 7.2 & 29.3 
\\
$10^{-0.5}$--$1$       & 8.6 & 3.2 & 1.5 & 8.9 & 18.3 
\\
$1$--$10^{0.5}$       & 2.1 & 0.80 & 1.3 & 2.2 & 6.4 
\\
$10^{0.5}$--$10$       & 0.09 & 0 & 0.49 & 0.40 & 0.98 
\\
\hline
TOTAL & 60.7 & 21.9 & 3.3 & 23.4 & 109.3
\\
\hline
\end{tabular}
\end{tabular}
\end{center}
\caption{
Spectrum of stable species from gluon and a tri-quark 
jets of 100 GeV. 
\label{tab2}}
\end{table}

\begin{table}
\begin{center}
\begin{tabular}{rl}
\begin{tabular}{r}
from q-jet\\
(10 GeV)
\\
\\
\\
\\
\\
from qq-jet
\\
(10 GeV)
\end{tabular}
&
\begin{tabular}{|c|c|c|c|c|c|}
\hline
{Energy [GeV]} & $\nu$ & $e$ & $p$ & $\gamma$ 
& TOTAL 
\\
\hline
$10^{-2}$--$10^{-1.5}$       & 5.5 & 1.9 & 0 & 0.91 & 8.3 
\\
$10^{-1.5}$--$10^{-1}$       & 5.2 & 1.8 & 0 & 2.0 & 9.0 
\\
$10^{-1}$--$10^{-0.5}$       & 2.9 & 1.2 & 0 & 1.7 & 5.9 
\\
$10^{-0.5}$--$1$       & 0.57 & 0.26 & 0.67 & 0.72 & 2.2 
\\
$1$--$10^{0.5}$       & 0 & 0 & 0.18 & 0.20 & 0.38 
\\
\hline
TOTAL & 14.2 & 5.2 & 0.85 & 5.5 & 25.8
\\
\hline
\hline
$10^{-2}$--$10^{-1.5}$       & 4.8 & 2.0 & 0 & 1.0 & 7.8 
\\
$10^{-1.5}$--$10^{-1}$       & 5.0 & 1.7 & 0 & 1.6 & 8.3 
\\
$10^{-1}$--$10^{-0.5}$       & 2.9 & 1.0 & 0 & 1.7 & 5.6 
\\
$10^{-0.5}$--$1$       & 0.85 & 0.34 & 0.59 & 0.74 & 2.5 
\\
$1$--$10^{0.5}$       & 0 & 0 & 0.20 & 0.14 & 0.34 
\\
\hline
TOTAL & 13.6 & 5.1 & 0.79 & 5.2 & 24.7
\\
\hline
\end{tabular}
\end{tabular}
\end{center}
\caption{
Spectrum of stable species from quark and diquark jets of 10 GeV. 
\label{tab3}}
\end{table}


\begin{thebibliography}{99}

\bibitem{ArkaniHamed:1998rs}
  N.~Arkani-Hamed, S.~Dimopoulos and G.~R.~Dvali,
  Phys.\ Lett.\  B {\bf 429} (1998) 263;
  I.~Antoniadis, N.~Arkani-Hamed, S.~Dimopoulos and G.~R.~Dvali,
  Phys.\ Lett.\  B {\bf 436} (1998) 257;
  L.~Randall and R.~Sundrum,
  Phys.\ Rev.\ Lett.\  {\bf 83} (1999) 3370.

\bibitem{Yao:2006px}
  W.~M.~Yao {\it et al.}  [Particle Data Group],
  J.\ Phys.\ G {\bf 33} (2006) 1.

\bibitem{Feng:2008ya}
  J.~L.~Feng and J.~Kumar,
  ``The WIMPless Miracle,''
  arXiv:0803.4196 [hep-ph].

\bibitem{Banks:1999gd}
  T.~Banks and W.~Fischler,
  ``A model for high energy scattering in quantum gravity,''
  arXiv:hep-th/9906038;
  R.~Emparan,
  Phys.\ Rev.\  D {\bf 64} (2001) 024025;
  S.~B.~Giddings and S.~D.~Thomas,
  Phys.\ Rev.\  D {\bf 65} (2002) 056010.
  D.~M.~Eardley and S.~B.~Giddings,
  Phys.\ Rev.\  D {\bf 66} (2002) 044011.

\bibitem{Dimopoulos:2001ib}
  S.~Dimopoulos and G.~L.~Landsberg,
  Phys.\ Rev.\ Lett.\  {\bf 87} (2001) 161602;
  for a review, see M.~Cavaglia,
  Int.\ J.\ Mod.\ Phys.\  A {\bf 18} (2003) 1843

\bibitem{Giudice:2001ce}
  R.~Emparan, M.~Masip and R.~Rattazzi,
  Phys.\ Rev.\  D {\bf 65} (2002) 064023;
  G.~F.~Giudice, R.~Rattazzi and J.~D.~Wells,
  Nucl.\ Phys.\  B {\bf 630} (2002) 293;
  J.~I.~Illana, M.~Masip and D.~Meloni,
  Phys.\ Rev.\ Lett.\  {\bf 93} (2004) 151102.

\bibitem{Corcella:2000bw}
  G.~Corcella {\it et al.},
  JHEP {\bf 0101} (2001) 010.

\bibitem{Battaner:2000ef}
  E.~Battaner and E.~Florido,
  Fund.\ Cosmic Phys.\  {\bf 21} (2000) 1
  [arXiv:astro-ph/0010475].

\bibitem{Illana:2005pu}
  J.~I.~Illana, M.~Masip and D.~Meloni,
  Phys.\ Rev.\  D {\bf 72} (2005) 024003
  [arXiv:hep-ph/0504234].

\bibitem{Sessolo:2008ex}
  E.~M.~Sessolo and D.~W.~McKay,
  arXiv:0803.3724 [hep-ph].

\bibitem{Pumplin:2002vw}
  J.~Pumplin, D.~R.~Stump, J.~Huston, H.~L.~Lai, P.~M.~Nadolsky and W.~K.~Tung,
  JHEP {\bf 0207} (2002) 012.

\bibitem{Hawking:1974rv}
  S.~W.~Hawking,
  Nature {\bf 248} (1974) 30;
  S.~W.~Hawking,
  Commun.\ Math.\ Phys.\  {\bf 43} (1975) 199
  [Erratum-ibid.\  {\bf 46} (1976) 206].

\bibitem{Draggiotis:2008jz}
  P.~Draggiotis, M.~Masip and I.~Mastromatteo,
  JCAP {\bf 07} (2008) 014.

\bibitem{Semikoz:2003wv}
  D.~V.~Semikoz and G.~Sigl,
  JCAP {\bf 0404} (2004) 003.

\bibitem{pamela} 
  M.~Boezio {\it et al.}  [PAMELA Collaboration],
  J.\ Phys.\ Conf.\ Ser.\  {\bf 110}, 062002 (2008);
  http://pamela.roma2.infn.it/

\bibitem{Gehrels:1999ri}
  N.~Gehrels and P.~Michelson,
  Astropart.\ Phys.\  {\bf 11} (1999) 277;
  http://www-glast.stanford.edu/

\bibitem{Abdo:2007ad}
  A.~A.~Abdo {\it et al.},
  Astrophys.\ J.\  {\bf 664} (2007) L91;
  http://www.lanl.gov/milagro/

\bibitem{Navarro:1995iw}
  J.~F.~Navarro, C.~S.~Frenk and S.~D.~M.~White,
  Astrophys.\ J.\  {\bf 462} (1996) 563.

\bibitem{Adriani:2008zr}
  O.~Adriani {\it et al.},
  ``Observation of an anomalous positron abundance in the cosmic radiation,''
  arXiv:0810.4995 [astro-ph].

\bibitem{Abdo:2008if}
  A.~A.~Abdo {\it et al.},
  ``A Measurement of the Spatial Distribution of Diffuse TeV Gamma Ray Emission
  from the Galactic Plane with Milagro,''
  arXiv:0805.0417 [astro-ph].

\bibitem{Atkins:2005wu}
  R.~W.~Atkins {\it et al.}  [The Milagro Collaboration],
  Phys.\ Rev.\ Lett.\  {\bf 95} (2005) 251103.

\bibitem{Prodanovic:2006bq}
  T.~Prodanovic, B.~D.~Fields and J.~F.~Beacom,
  Astropart.\ Phys.\  {\bf 27} (2007) 10.

\bibitem{Casanova:2007cf}
  S.~Casanova and B.~L.~Dingus,
  Astropart.\ Phys.\  {\bf 29} (2008) 63.

\bibitem{Hunger:1997we}
  S.~D.~Hunter {\it et al.},
  Astrophys.\ J.\  {\bf 481} (1997) 205.

\bibitem{Baughman:2007ck}
  B.~M.~Baughman, W.~B.~Atwood, R.~P.~Johnson, T.~A.~Porter and M.~Ziegler,
  ``A Fresh Look at Diffuse Gamma-ray Emission from the Inner Galaxy,''
  arXiv:0706.0503 [astro-ph].

\bibitem{Barrau:2005zb}
  A.~Barrau, C.~Feron and J.~Grain,
  Astrophys.\ J.\  {\bf 630} (2005) 1015.

\bibitem{Bringmann:2007nk}
  T.~Bringmann, L.~Bergstrom and J.~Edsjo,
  JHEP {\bf 0801} (2008) 049.

\bibitem{Cirelli:2008jk}
  M.~Cirelli and A.~Strumia,
  ``Minimal Dark Matter predictions and the PAMELA positron excess,''
  arXiv:0808.3867 [astro-ph].

\bibitem{Barger:2008su}
  V.~Barger, W.~Y.~Keung, D.~Marfatia and G.~Shaughnessy,
  ``PAMELA and dark matter,''
  arXiv:0809.0162 [hep-ph].

\bibitem{Cholis:2008hb}
  I.~Cholis, L.~Goodenough, D.~Hooper, M.~Simet and N.~Weiner,
  ``High Energy Positrons From Annihilating Dark Matter,''
  arXiv:0809.1683 [hep-ph].

\bibitem{Hooper:2008kg}
  D.~Hooper, P.~Blasi and P.~D.~Serpico,
  ``Pulsars as the Sources of High Energy Cosmic Ray Positrons,''
  arXiv:0810.1527 [astro-ph].

 
\end{thebibliography}
\end{document}